\documentclass[12pt]{iopart}
\usepackage{iopams}
\usepackage{graphicx}
\usepackage{epsfig}
\usepackage{graphics}
\usepackage{setspace}
\usepackage{vmargin}
\begin{document}

\title[ Rheology of Poly(vinyl-acetate)
 Monolayers]{Shear and Compression Viscoelasticity in Polymer
 Monolayers}

\author{Toby A. M. Ferenczi \dag\ and Pietro Cicuta \dag
\footnote[3]{To whom correspondence should be addressed
(pc245@cam.ac.uk)} }

\address{\dag\ Cavendish Laboratory, University of Cambridge, Madingley Road, Cambridge Cb3 0HE, U.K.}


\begin{abstract}
Poly-vinlyacetate (PVAc)  forms very stable and reproducible
monolayers on the surface of water,  a model system to understand
polymer physics on two dimensions. A recently introduced technique
is applied here to to study  viscoelasticity of PVAc monolayers.
The method is based on measurement of surface tension in two
orthogonal directions during anisotropic deformation. Compression
and shear moduli are explored over a very large concentration
range, highlighting a series of four different regimes. At low
concentration the polymers are in a dilute gas. Above the overlap
concentration $\Gamma^\ast$ there is a fluid semi-dilute region,
where the monolayer properties are described by scaling laws. At a
threshold concentration $\Gamma^{\ast\ast}$, a decrease in the
gradient of pressure with concentration is observed, and we argue
that there is still a large fraction of free area on the surface.
Compressing further, we then identify close packing as the point
where  the pressure gradient  rises sharply  and a shear modulus
emerges. This is interpreted as a transition to a soft-solid
  due to the kinetic arrest of close-packed
monomers. The reological properties of PVAc above
$\Gamma^{\ast\ast}$  have not been studied previously. Discussion
includes possible explanations for the observed behaviour in terms
of both equilibrium and non-equilibrium conditions, and the
relation to microscopic chain properties. Temperature dependent
effects around $\Gamma^{\ast\ast}$ are also observed and
described.

\end{abstract}

\section{Introduction}

Polymer physics has evolved over the last 50 years into a highly
successful and sophisticated set of
theories~\cite{deGennes79,edwards86}. A significant area of
uncertainty remains in the application of some of these ideas to
polymers confined to two dimensions (2D). This is partly due to
the greater experimental challenge in establishing well controlled
two dimensional systems. One approach is to spread polymers on the
surface of a liquid, to form Langmuir films. Recent developments
in surface rheometry \cite{fuller99,monroy98,monroy03,cicuta04}
 are now enabling experiments that probe the viscoelastic
properties of 2D polymer solutions as a function of concentration
and frequency, in analogy with similar classical experiments in
bulk solutions. Some  three dimensional behavior has a direct
correspondence in two dimensions, for example the existence of a
semi-dilute concentration regime where the excluded volume effect
is progressively screened~\cite{rondelez80}. Recent experiments
indicate that other behavior, in particular relating to chain
dynamics, is unique to two dimensions. For example for a large set
of different monolayers the compressional dynamics is described by
a timescale that is not related to the classical reptation
mechanism~\cite{cicuta04}.

 In
this work a very well studied system is chosen: monolayers of
poly-vinlyacetate (PVAc) on the surface of
water~\cite{rondelez80,richards99,yu89,kim89,monroy98,monroy03}.
The equilibrium and dynamical properties of this system are well
known at low concentrations. In this work  measurements are
extended to very high concentration, where surprisingly we observe
the development of a finite shear modulus. The presence of
 a glass transition at low temperatures has been suggested for
PVAc layers~\cite{monroy00b} and this is also re-examined here.


\section{Theoretical Background\label{theoreticalbackground}}

\subsection{Viscoelasticity of Monolayers\label{viscoelasticity of monolayers}}

In a Langmuir trough experiment the  surface tension is measured
as a function of available surface area. For an insoluble
monolayer the surface concentration is related inversely to the
area $A$, $\Gamma=M/A$, where $M$ is the mass on the surface.
Surface pressure is identified as the resultant drop in surface
tension as concentration increases, from the value $\gamma_0$ of
the clean interface, $\Pi_{eq}=\gamma_0-\gamma$.

In general, the response to an arbitrary deformation can be
characterised by two independent moduli: compression,
$\varepsilon$ and shear, $G$. For isotropic and quasi-static
compressions  the `equilibrium modulus' $\varepsilon_{eq}$ is
probed, whereas at finite compression strain rates, a viscosity
$\eta_d$ would be observed. $\varepsilon_{eq}$ is an elastic
(storage) component proportional to the derivative of pressure
with area, and $\eta_d$ is due to dissipation from frictional
resistance to the flow:
\begin{equation}
\varepsilon_{eq}=\Gamma\left(
\frac{\partial\Pi_{eq}}{\partial\Gamma} \right)_T
\,\,\,\,\textrm{and}\,\,\,\,
\eta_d=A\frac{\Pi-\Pi_{eq}}{\frac{d}{dt}A}. \label{definition of
equilibrium mod}
  \end{equation}
  The
complex shear modulus is defined as the ratio of shear stress
response to an induced strain (at constant area). As usual in
linear viscoelasticity,   the
 complex dynamic
 moduli for compression ($\varepsilon^\ast$) or shear ($G^\ast$)
 can be
measured  following oscillatory deformations,
\begin{equation}
\varepsilon^\ast(\omega)=\varepsilon'(\omega)+i\varepsilon''(\omega);
\,\,\, \,\,G^\ast(\omega)=G'(\omega)+iG''(\omega),
\label{complexeps}
\end{equation}
where $\omega$ is the frequency of oscillation and
$\varepsilon''(\omega)=\omega\eta_d(\omega)$. In
eq.~\ref{complexeps} the real and imaginary parts describe the
in-phase (elastic) and the out of phase (dissipative) components
of the response.

\begin{figure}[t]
\centering
 \epsfig{file=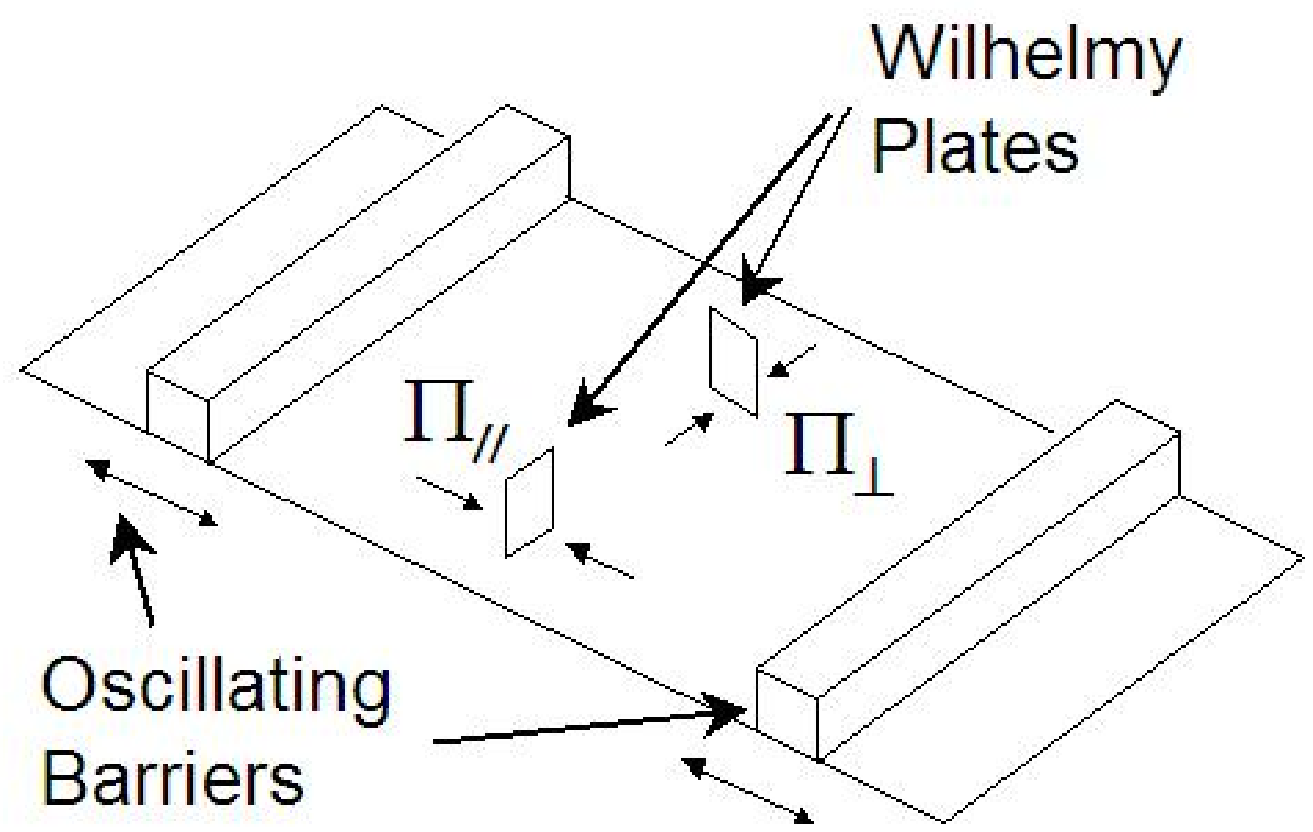, width = 5cm} \,\,\, \epsfig{file=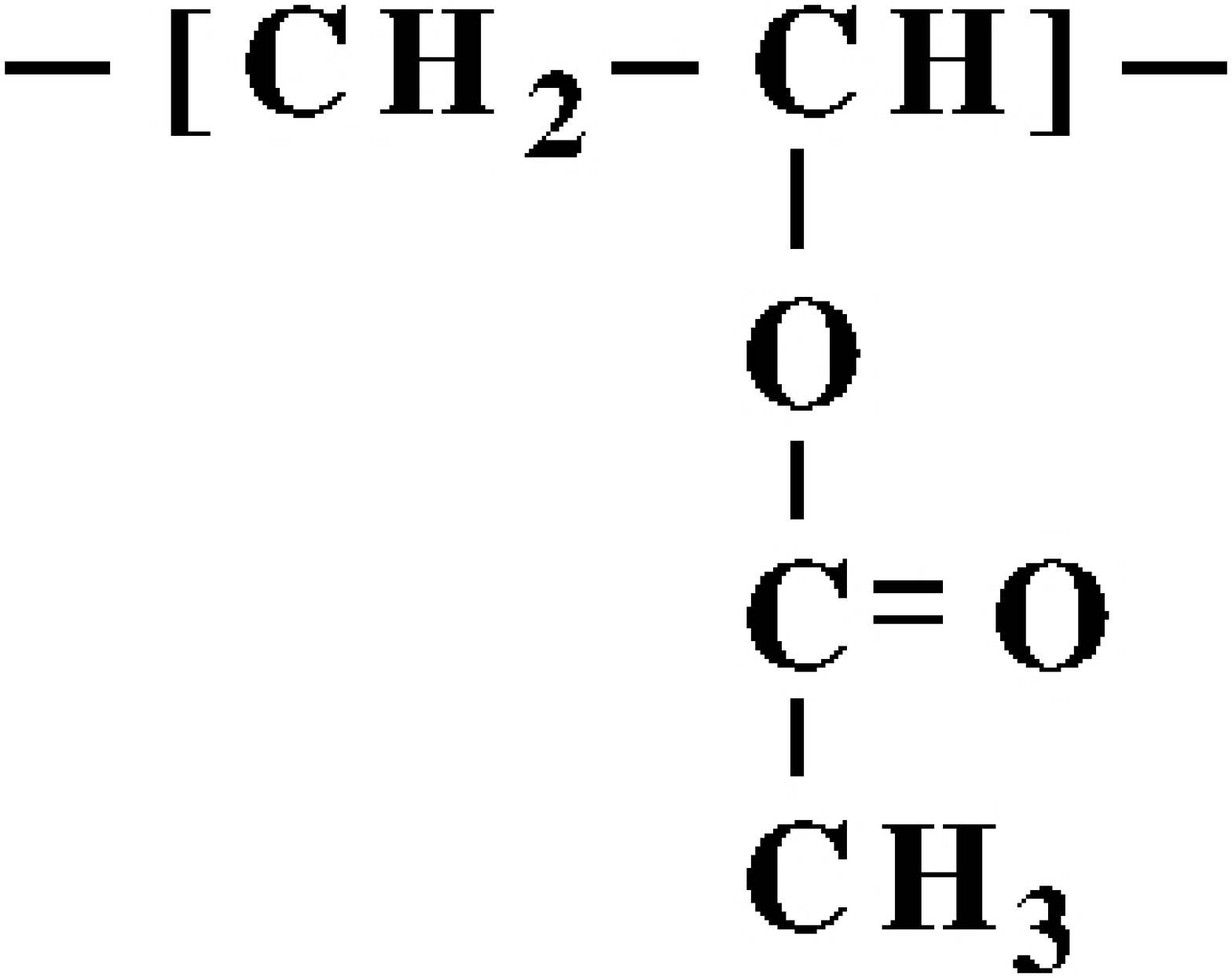,width=3cm}
 \caption{(a)~Schematic diagram of Langmuir trough setup for anisotropic measurements of surface pressure.
 (b)~Chemical diagram of vinyl acetate monomer~\cite{bandrup99}. \label{trough}}
\end{figure}

As  recently described in~\cite{cicuta04b}, it possible in surface
monolayers to determine both $\varepsilon^\ast$ and $G^\ast$ from
the full stress response $\Pi(t)$. This is because under uniaxial
compression of a Langmuir film
 \textit{both} compression and shear deformation are exerted, although
contribution from shear is often small and  treated as negligible.
Petkov \textit{et al.}~\cite{petkov00} were the first to show  a
directional anisotropy in surface pressure measurements using two
Wilhelmy plates arranged in orthogonal directions, see
Fig.~\ref{trough}. This anisotropy is directly dependent on the
shear modulus.
It was shown that for sinusoidal deformations of the form
$\delta{A(t)}/A_0=\Delta{A}/A_0\cos{\omega{t}}$, the pressure
response can be expressed as
\begin{eqnarray}
\Pi_{\parallel}-\Pi_0=\delta\Pi_{\parallel}(t)=\frac{\Delta{A}}{A_0}\left[(\varepsilon'+G')\cos{\omega{t}}+(\varepsilon''+G'')\sin{\omega{t}}\right]
\nonumber \\
\Pi_{\perp}-\Pi_0=\delta\Pi_{\perp}(t)=\frac{\Delta{A}}{A_0}\left[(\varepsilon'-G')\cos{\omega{t}}+(\varepsilon''-G'')\sin{\omega{t}}\right].
\end{eqnarray}

\subsection{Concentration regimes for polymers in monolayers\label{chain theory}}
Although the separation of concentration into three regions
(dilute, semi-dilute and concentrated) has been proposed before,
it is worth summarizing here some of the key ideas, because they
underpin the discussion of the different dynamical response
regimes studied in this paper. For isolated polymer chains the
mean end-to-end distance is well known, $R\simeq{aN^{\nu}}$, given
that $\nu$ is the Flory exponent~\cite{flory53}, $a$ is the
monomer size and $N$ the number of monomers per chain. This
expression for $R$ would be an equality if $a$ were replaced by
the  statistical Kuhn length $b$, and $N$ by $aN/b$. For very
flexible polymers like PVAc the two lengths $a$ and $b$ are likely
to be very similar. In the case of two dimensions and for excluded
volume type interactions, theoretical predictions for neutral
polymers give $\nu=0.75$~\cite{deGennes79,rondelez80}. Such a
chain is said to be in `good' solvent conditions and obeys a self
avoiding walk conformation.

If the average separation between chains is greater than $R$ the
system can be thought of as a two dimensional gas, resulting in a
linear relation between surface pressure and surface
concentration.  Chains are no longer isolated above the overlap
concentration, $\Gamma^\ast$, defined as the concentration where
overall surface concentration is the same as that within each
unperturbed chain. The dependence of $\Gamma^\ast$ on $N$ follows
from this definition,
\begin{equation}
\Gamma^{\ast}\sim\frac{N\,\,{\textmd{[monomer mass]}}}{{R}^2}\sim
N^{(1- 2 \nu)}.\label{gammastar1}
\end{equation}

 The behaviour above
$\Gamma^{\ast}$ is known as the semi-dilute regime and detailed
description may be found in refs.~\cite{deGennes79,edwards86}.
Briefly, the presence of surrounding
 chains results in the progressive screening of intra-chain repulsive
 interactions,
until eventually at high monomer density (a melt) the  behaviour
of the chain becomes ideal. A characteristic length, $\xi$, can be
introduced: below $\xi$ the chain does not interact with other
chains and thus still obeys a self avoiding walk; above $\xi$ the
chain can be seen as a succession of `blobs' following an ideal
random walk. At $\Gamma^{\ast}$, $\xi\sim{R}$ and, as the chains
are compressed, the characteristic length decreases rapidly.
Scaling laws are obtained for the equilibrium pressure and
compression modulus of a polymer film by assuming that the
equilibrium properties in the semi-dilute regime only depend on
$\xi$ and not on $N$:
\begin{equation}
\Pi_{eq}  \simeq
\frac{k_BT}{{R}^2}\left(\frac{\Gamma}{\Gamma^\ast}\right)^{y_{eq}}
\,\,\,\,\,\textrm{and} \,\,\,\,\,
\varepsilon_{eq}  =
y_{eq}\,\Pi_{eq},\,\,\,\,\,\textrm{with}\,\,\,y_{eq} =
\frac{2\nu}{(2\nu-1)}. \label{scalinglawforepsillon}
\end{equation}
    Polymer monolayers in the semidilute regime are
fluid, and have negligible shear elasticity and viscosity. Their
compressional dynamics has been the focus of recent investigations
that have shown  scaling of the compressional
viscosity~\cite{cicuta04}, a  particularly surprising result
because it indicates a relaxation mechanism specific to two
dimensional layers.

The semidilute regime ends at a concentration $\Gamma^{\ast\ast}$
when the correlation length $\xi$ becomes of the order of the
monomer size. To be more precise in defining $\Gamma^{\ast\ast}$,
the distinction should be made between three lengths: the monomer
size $a$, the statistical length (Kuhn length) $b$, and $R_{sw}$.
The distinction between $a$ and $b$ also matters in determining
exactly $\Gamma^\ast$, and was introduced  above. The   length
scale $R_{sw}$ becomes relevant if there are repulsive
interactions between monomers described in terms of a Flory-type
positive second-virial coefficient,
$v_2$~\cite{edwards86,andelman03}. Then it is found that swollen
behaviour (i.e. $R\sim{N^{0.75}}$) is only realised above a
minimal monomer number $N_{sw}$, below which chain statistics are
unaffected by the interaction.  $R_{sw}$ is expected to be of the
order of a few segment lengths $a$~\cite{andelman03}.  The
semi-dilute regime will end when the `characteristic length' $\xi$
reduces to the largest of the length scales discussed here, that
is $R_{sw}$. At that point the entire system becomes statistically
ideal. There are currently no experiments
 to distinguish precisely which of $a, \, b $ or
$R_{sw}$ are limiting the semidilute regime, and different
polymers may not be limited by the same length. The key point in
considering $\Gamma^{\ast\ast}$ is that the area fraction
$\Phi^{\ast\ast}$ actually covered by monomers at this
concentration can  be quite low, somewhere between 20\% to 35\%.
These very rough estimates are based either on $R_{sw}$ being
between $2a$ and $3a$, or considering that $\xi$ reduces all the
way down to $a$, but observing that the monomer area is roughly
$a^2/3$ which is plausible given the monomer structure, see
fig.~\ref{trough}(b). Measurements discussed below of the ratio
$\Gamma^{\ast\ast}/\Gamma^\ast$ support this estimate.

\section{Experimental Methods\label{method}}
Experimental methods are very similar to those described in
\cite{cicuta04b} for measurements on protein layers, so only the
most important facts are summarized here. The monolayer is
contained within a Langmuir trough of total area 530cm$^2$ and
width 20cm
 with two symmetrical barriers (mod. 610, Nima Technology, UK).
Surface pressures $\Pi_\parallel$ and $\Pi_\perp$ are determined
using two filter paper Wilhelmy plates, positioned at the center
of the trough, one parallel and the other perpendicular to the
compression direction. Polymer solution, typically $60\mu{l}$ of a
0.1mg/ml solution in tetrahydrofuran (analytical grade, Fisher
Scientific) is spread onto an ultrapure water subphase  in a
dropwise fashion using a microsyringe.  After spreading, the layer
is left for at least 30 minutes to allow the solvent to evaporate
and for the layer to reach equilibrium. The PVAc used throughout
this investigation has molecular weight
 $M_w=170,000$g/mol (Acros organics), except for one experiment
 reported below where $M_w=17,000$g/mol is used (American Polymer
Standards Corporation). Temperature of the subphase is controlled
via  water circulating under the trough.

To measure viscoelasticity the area is changed by oscillatory
motion of the barriers, keeping   the fractional amplitude of
oscillation  constant at ${\Delta}A/A_0{\sim}2\%$. The surface
pressure response is recorded as a function of time by both
Wilhelmy plates. Maximum accuracy is achieved by  using the same
sensor, and repeating each experiment spreading  identical layers
 with the sensor's plate in each orientation.  This
is done at low concentrations, where any viscous effect is
expected to be small. Two different sensors are used
simultaneously as depicted in fig.~\ref{trough} when the effect of
the compressional viscosity and shear modulus are very large, as
at high concentrations (Fig.~\ref{beautiful_conc}). Data is
collected every 0.1s, and about ten oscillations are made at each
pressure, at two barrier speeds corresponding to periods of
roughly 10s and 24s. Consideration of the propagation time of
compression waves must be taken into account, as described in
ref.~\cite{cicuta04b}.


\section{Results and discussion\label{results}}
\subsection{Dilute to Semi-Dilute and Concentrated regimes}

\begin{figure}[h]
\centering
 \epsfig{file=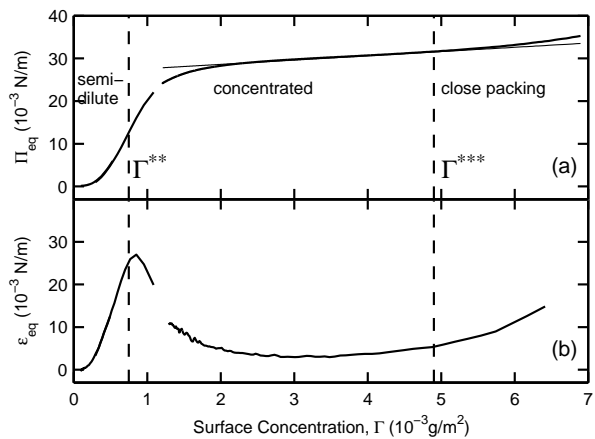, width = 8cm}
 \caption{(a) Surface pressure, $\Pi_{eq}$  and (b)~equilibrium compression modulus $\varepsilon_{eq}$
 as a function of surface concentration over the entire concentration range at $6^oC$.
 The straight  line interpolating the data between 2 and 5mg/m$^2$ has a gradient of 1.
 The vertical lines identify the end of the semi-dilute regime  and the deviation from a
 linear gradient in the pressure.\label{press_vs_conc}}
\end{figure}

Figures~\ref{press_vs_conc}(a) and~\ref{press_vs_conc}(b) are
isotherms showing equilibrium surface pressure $\Pi_{eq}$ and
compression modulus $\varepsilon_{eq}$. The curves are independent
of the orientation of the Wilhelmy plates.  There is a power-law
region extending up to $\Gamma^{\ast\ast} = 0.75$mg/m$^2$, where
the exponent  $y_{eq} = 2.8 \pm 0.2$, corresponding to a Flory
exponent of $\nu = 0.78 \pm 0.03$. This is in agreement with
`good' solvent predictions and other studies of PVAc
\cite{monroy00b,kim89,rondelez80,monroy03,cicuta04}.
The only direct indication of  the point of overlap is the change
in slope from $1$ to $y_{eq}$ in a logarithmic plot of surface
pressure with concentration. For high molecular weight
$M_w=170000,\, N=1977$ and the dilute regime ends at a very low
concentration. Therefore the linear relation between pressure and
concentration is at very low pressures, below the experimental
resolution, in agreement with~\cite{rondelez80}. Isotherm
measurements with a much lower molecular weight $M_w=17000$
($N\simeq 200$) (not shown) do display a clear dilute-semidilute
transition, at $\Gamma^{\ast}_{17000}=0.17\pm 0.01$mg/m$^2$ and
$\Pi^{\ast}_{17000}=0.45\pm 0.02$mN/m. This is in good agreement
with the value of $\Gamma^{\ast}_{17000}=0.19$ that is obtained
using eq.~\ref{gammastar1} and estimating $R$ from the known bond
length of $a=0.23$nm~\cite{bandrup99}. However these measurements
of $\Gamma^{\ast}$ are very delicate, because of the very low
pressures involved and in particular the very high compressibility
of the gas phase, that causes extremely long equilibration times
(even hours) for the concentration across the surface.
 Indeed, the same overlap values as seen
 for   $M_w=17000$ have been reported recently for   $M_w=90000$~\cite{monroy03}.
  An uncontroversial determination of
 $\Gamma^{\ast}$ can only come from measurements on a series of
 molecular weights.   For $M_w=17000$ a value of  $\Gamma= 0.80$ is obtained for the peak position of
 the dilational modulus,
 the same as in Figure~\ref{press_vs_conc}(b). This means that the
$M_w=17000$
 layer has been compressed by a factor of around~4 between
 $\Gamma^{\ast}$ and  $\Gamma^{\ast\ast}$. The packing fraction $\Phi$ of monomers can be estimated assuming
 each monomer occupies an area $a^2$. Then  the packing fraction
 for $M_w=17000$ at  $\Gamma^{\ast}$ is $\Phi^{\ast}\simeq 0.07$, and at
 $\Gamma^{\ast\ast}$ $\Phi^{\ast\ast}\simeq 0.28$. This confirms
 the previous argument on the possibility of considerable free
 space at  $\Gamma^{\ast\ast}$.


\begin{figure}[h]
\centering
 \epsfig{file=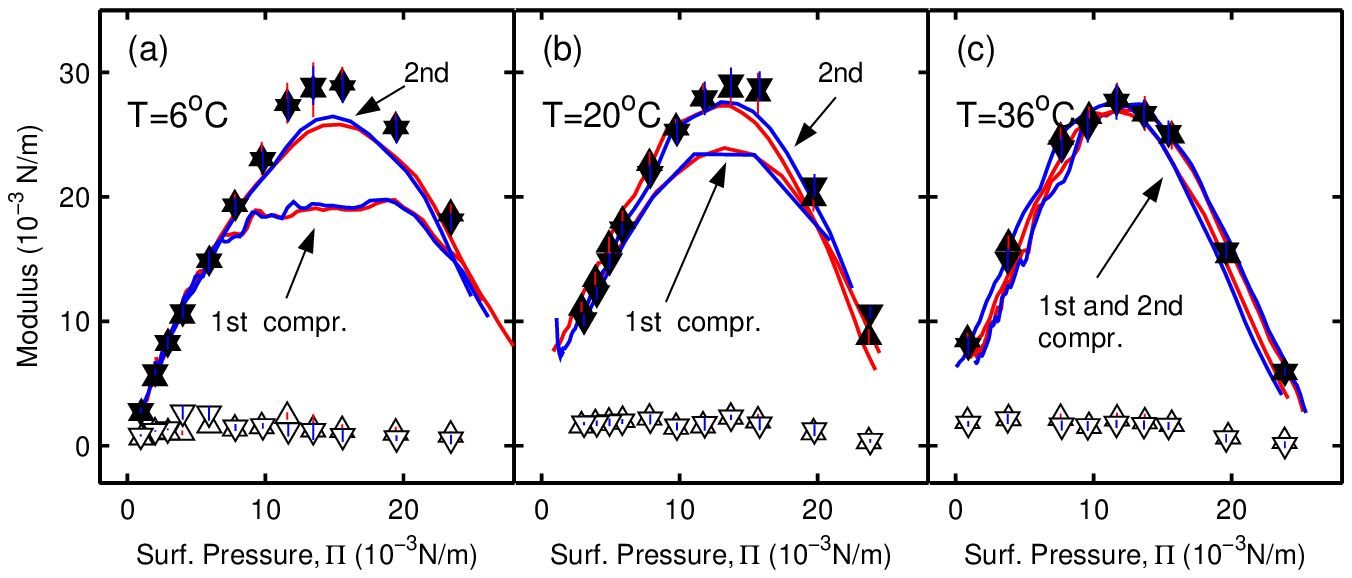, width = 10cm}\\  \epsfig{file=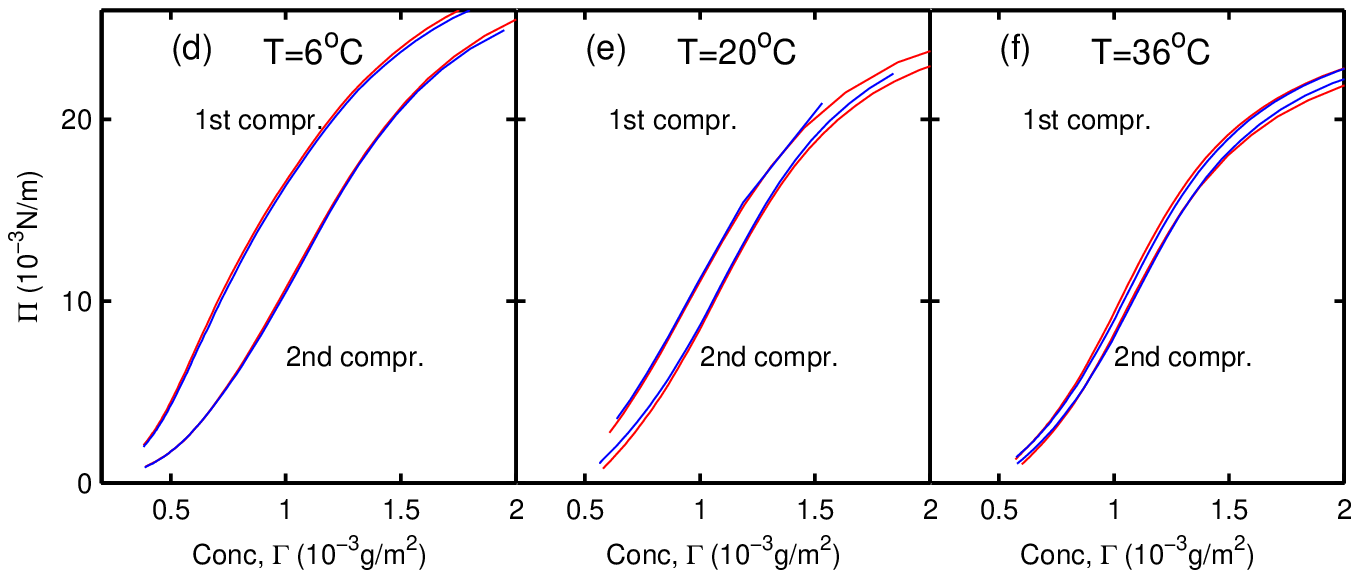, width = 10cm}
 \caption{In (a), (b) and (c) lines are $\varepsilon_{eq}$ recorded in
  perpendicular orientations, at different temperatures.
 Differences between first and second compressions are very marked at low temperature. Symbols correspond to
  measurements of dynamic moduli: ($\blacktriangle$) $\varepsilon'+G'$,  ($\blacktriangledown$) $\varepsilon'-G'$,
  ($\vartriangle$) $\varepsilon''+G''$ and  ($\triangledown$) $\varepsilon''-G''$. The concentration loss effect described in the text
  between first and second compression
 isotherms is shown in (d), (e) and (f).
  $\Pi_{\parallel}$ and $\Pi_{\perp}$ are indistinguishable in this range.
\label{cold}}
\end{figure}

Above  $\Gamma^{\ast\ast}$, Figure~\ref{press_vs_conc}(a) shows an
interesting linear dependence of the surface pressure on the
concentration. In this regime, that we call concentrated,  the
concentration increases by a factor of roughly 6 from
$0.75$mg/m$^2$ to $4.8$mg/m$^2$. The  linear pressure dependence
could be due to the entropic cost of compressing what is
essentially a gas of monomers rather than swollen regions. This
picture assumes that the system remains in equilibrium throughout.
Further
investigation of the pressure divergence as the area per monomer
decreases are needed to check this hypothesis.

\subsection{Temperature effects} The measurements in figures~\ref{cold}(a), (b) and (c)
correspond to the semi-dilute and beginning of the  concentrated
regimes. A surprising result is that at low temperatures the first
compression isotherm differs from a second compression exerted
after re-expansion of the same monolayer. Further compressions are
identical to the second. This effect is most evident at the lowest
temperature. Looking at figure~\ref{cold}(d) it would appear that
there has been a loss of concentration after the first
compression. We think that at low temperatures some of the polymer
conformations  with long timescales become frozen out due to
stearic or `jamming' effects. The system then enters a regime
where only short timescale configurations are taking place which
causes a reduction in the free energy cost of compression, leading
to the observed drop in pressure gradient. Upon re-expansion these
conformations remain `frozen', thus reducing the effective density
of the layer.  The experiment reported here seems another
manifestation of the temperature dependence of the thermal
expansion coefficient reported in~\cite{monroy00b}. Note that for
compressions after the first, the semi-dilute regime shows very
similar behavior at all temperatures. In all experiments in this
concentration range
 $\varepsilon'+G'$ and $\varepsilon'-G'$ are the same,
which implies close to zero shear elasticity in the semidilute
region. Well below ${\Gamma^{\ast\ast}}$ $\varepsilon'$ and
$\varepsilon_{eq}$ are indistinguishable. Close to
${\Gamma^{\ast\ast}}$ the complex compressional elasticity,
$\varepsilon'$ becomes consistently greater than the its
equilibrium counterpart, $\varepsilon_{eq}$. The compression
dynamic modulus remains in the power law regime until slightly
higher pressures, which may be due to the  small dynamical
excitations allowing the polymers to explore more chain
configurations (as described in \cite{cicuta04b}). The complex
viscosities are finite but small ($G''$ always $\leq 2$mN/m) at
this frequency and in this concentration range.  Recent studies of
PVAc Langmuir films \cite{monroy03,cicuta04} have independently
shown and explained with two different models the presence of
scaling laws for compressional viscosity  where the exponent is
twice that for compressional elasticity , i.e.
$\varepsilon''\sim{\Gamma^{2y_{eq}}}$. This implies that
compressional viscosity should follow a quadratic function of
pressure, but the data presented here is unable to resolve this
trend.

\subsection{Close Packing}
The linear dependence of pressure on concentration, as shown by
the dashed line of fig.~\ref{press_vs_conc}(a), lasts up to a
concentration  $\Gamma^{\ast\ast\ast}= 4.8$mg/m$^2$. We think that
this concentration   is very close to close packing of monomers on
the monolayer. Even higher values  (not shown here) of the
pressure can be reached, however  they are not stable over time.
30~hours after a compression to $\Pi \simeq 36$mN/m, the pressure
had equilibrated to around $31.6$mN/m, the value at
$\Gamma^{\ast\ast\ast}$.
\begin{figure}[t]
\centering
 \epsfig{file=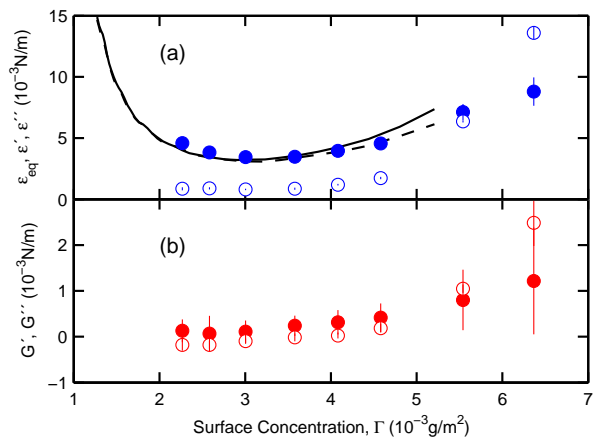, width=8cm}
 \caption{(a) Compression and (b) Shear moduli concentration, at very high concentration and T=$6^oC$.
  Solid and dashed
  lines show
 $\varepsilon_{eq}$ in parallel and perpendicular orientations respectively.
 Solid and empty symbols are the elastic and viscous components. \label{beautiful_conc}}
\end{figure}
Figure~\ref{beautiful_conc}  shows that below $\Gamma= 3$mg/m$^2$
the shear modulus, $G'$ remains zero and the equilibrium elastic
modulus $\varepsilon_{eq}$ is indistinguishable from the dynamic
compressional modulus $\varepsilon'$. At a critical concentration,
$\varepsilon'$ increases sharply with pressure and eventually
exceeds $\varepsilon_{eq}$. The viscous modulus $\varepsilon''$
also increases sharply after  $\Gamma= 4$mg/m$^2$ and becomes
greater than the elastic component. Both $G'$ and $G''$ become
non-zero approaching close packing, although neither becomes as
large as the compressional components. We further remark that the
viscous shear component exceeds the elastic component at high
concentration. Detailed analysis of these trends is premature.

It is probable that the shear modulus  approaching close packing
arises by a process of dynamical arrest due to crowding rather
than by formation of a network structure. It has been shown from
both simulation~\cite{santen00} and experiment~\cite{cicuta03}
that systems of colloidal monolayers undergo kinematic arrest at
surface fractions of $\Phi\simeq0.8$. The effect is essentially
caused by the caging of a particle by its neighbours due to hard
core repulsion interactions. At high surface fractions, the
behaviour of vinyl acetate monomers may be dominated by their hard
core interactions as in colloidal systems. Some simulations have
suggested that two dimensional systems of polymer chains may form
interpenetrated, entangled networks at high concentration
\cite{ostrovsky97} whereas several other
simulations~\cite{kremer90} and experiments~\cite{radler99} have
pointed out that chains confined to two dimensions remain as
segregated disks, as also suggested in \cite{deGennes79}.  Our
results appear to support the latter case because it is unlikely
that solid behaviour would emerge only at $\Phi\sim 1$ if it was
due entanglements.

The instability in the layer above $\Gamma^{\ast\ast\ast}$
indicates that there is likely to be some  collapse into the
subphase at this concentration, but because such an effect is
observed only at extreme concentrations we do not consider it
necessary to explain the observed behavior at lower concentrations
in terms of  out-of-plane polymer rearrangement or multilayer
formation.


\section{Conclusions\label{conclusions}}
The set of  experiments reported here characterize the
viscoelastic properties of poly(vinyl-acetate) monolayers over a
concentration range of several orders of magnitude. Within this
range we have identified four types of behaviour. Specifically,
these are the dilute, semi-dilute, concentrated  and close-packed
regimes, separated by the transition concentrations,
$\Gamma^{\ast}$, $\Gamma^{\ast\ast}$ and $\Gamma^{\ast\ast\ast}$.
The dilute and semi-dilute regimes have previously been well
defined and our results are in agreement with these descriptions.
At higher concentration, the behaviour has been subject to far
less scrutiny, due in part to experimental limitations. We have
observed a region of linear pressure gradient that we argue is
caused by the entropic cost of compressing a gas of monomers. This
is followed by a region of close-packed behaviour resulting in the
formation of a soft-solid, as evidenced by the emergence of a
shear modulus. Temperature dependence of the compression modulus
has been observed and interpreted as evidence of non-equilibrium
effects. Using existing polymer theory the importance of
microscopic length scales associated with the polymer chain has
been stressed. This implies a degree of universality in the
description of neutral homopolymers which presents interesting
opportunities for further study.\\

\ack  P.C. thanks the Oppenheimer Fund for financial support.\\

 \bibliographystyle{unsrt-notitle}
      \bibliography{bibdatav14}

\begin{thebibliography}{10}

\bibitem{deGennes79}
P.-G. de~Gennes.
\newblock {\em Scaling Concepts in Polymer Physics}.
\newblock Cornell University Press, Ithaca, 1979.

\bibitem{edwards86}
M.~Doi and S.~F. Edwards.
\newblock {\em The Theory of Polymer Dynamics}.
\newblock Oxford University Press, New York, 1986.

\bibitem{fuller99}
C.~F. Brooks, G.~G. Fuller, C.~W. Curtis, and C.~R. Robertson.
\newblock {\em Langmuir}, 15:2450, 1999.

\bibitem{monroy98}
F.~Monroy, F.~Ortega, and R.~G. Rubio.
\newblock {\em Phys. Rev. E}, 58:7629, 1998.

\bibitem{monroy03}
F.~Monroy, H.M.Hilles, F.~Ortega, and R.~G. Rubio.
\newblock {\em Phys. Rev. Lett.}, 91:268302, 2003.

\bibitem{cicuta04}
P.~Cicuta and I.~Hopkinson.
\newblock {\em Europhys. Lett.}, 68:65, 2004.

\bibitem{rondelez80}
R.~Vilanove and F.~Rondelez.
\newblock {\em Phys. Rev. Lett.}, 45:1502, 1980.

\bibitem{richards99}
R.~A.~L. Jones and R.~W. Richards.
\newblock {\em Polymers at Surfaces and Interfaces}.
\newblock Cambridge Univ. Press, Cambridge (U.K.), 1999.

\bibitem{yu89}
K.-H. Yoo and H.~Yu.
\newblock {\em Macromolecules}, 22:4019, 1989.

\bibitem{kim89}
B.~B. Sauer, H.~Yu, M~Yazdanian, G~Zografi, and M.W. Kim.
\newblock {\em Macromolecules}, 22:2332, 1989.

\bibitem{monroy00b}
F.~Monroy, F.~Ortega, and R.~G. Rubio.
\newblock {\em Eur. Phys. J. B}, 13:745, 2000.

\bibitem{cicuta04b}
P.~Cicuta and E.M. Terentjev.
\newblock {\em European Phys. J. E}, 16:147, 2005.

\bibitem{petkov00}
J.~T. Petkov, T.~D. Gurkov, B.~E. Campbell, and R.~P. Borwankar.
\newblock {\em Langmuir}, 16:3703, 2000.

\bibitem{bandrup99}
In J.~Bandrup, E.~H. Immergut, and E.~A. Grulke, editors, {\em Polymer
  Handbook}, New York, 1999. Wiley.

\bibitem{flory53}
P.J. Flory.
\newblock {\em Principles of Polymer Chemistry}.
\newblock Cornell University Press, Ithaca, 1953.

\bibitem{andelman03}
R.~R. Netz and D.~Andelman.
\newblock {\em Physics Reports}, 380:1--95, 2003.

\bibitem{santen00}
L.~Santen and W.~Krauth.
\newblock {\em Nature}, 405:550, 2000.

\bibitem{cicuta03}
P.~Cicuta, E.~J. Stancik, and G.~G. Fuller.
\newblock {\em Phys. Rev. Lett.}, 90:236101, 2003.

\bibitem{ostrovsky97}
B.~Ostrovsky, M.~Smith, and Y.~Bar-Yam.
\newblock {\em Int. J. Mod. Phys. B}, 8:931, 1997.

\bibitem{kremer90}
I.~Carmesin and K.~Kremer.
\newblock {\em J. Phys. France}, 51:915, 1990.

\bibitem{radler99}
B.~Maier and J.~O. R{\"{a}}dler.
\newblock {\em Phys. Rev. Lett.}, 82:1911, 1999.

\end{thebibliography}

\end{document}